\documentclass[manuscript]{acmart}

\usepackage{multirow}
\usepackage{diagbox}
\usepackage{graphicx}
\usepackage{ctable}

\AtBeginDocument{%
  }

\setcopyright{acmlicensed}
\copyrightyear{2018}
\acmYear{2018}
\acmDOI{XXXXXXX.XXXXXXX}

\acmConference[Conference acronym 'XX]{Make sure to enter the correct
  conference title from your rights confirmation emai}{June 03--05,
  2018}{Woodstock, NY}
\acmISBN{978-1-4503-XXXX-X/18/06}

\begin{document}

\title[Digital Reading Materials and Religious Practice]{Digital Reading Materials and Religious Practice: Usage in Personal Study and Religious Education Courses}

\author{Teancum Price}
\affiliation{%
  \institution{Utah State University}
  \city{Logan}
  \state{Utah}
  \country{USA}}
\email{teancum.price.dev@gmail.com}

\author{Musa Blake}
\affiliation{%
  \institution{Abilene Christian University}
  \city{Abilene}
  \state{Texas}
  \country{USA}}
\email{mbb23c@acu.edu}

\author{James Prather}
\affiliation{%
  \institution{Abilene Christian University}
  \city{Abilene}
  \state{Texas}
  \country{USA}}
\email{james.prather@acu.edu}

\author{Seth Poulsen}
\affiliation{%
  \institution{Utah State University}
  \city{Logan}
  \state{Utah}
  \country{USA}}
\email{seth.poulsen@usu.edu}

\renewcommand{\shortauthors}{Price et al.}

\newcommand{\churchname}{The Church of Jesus Christ of Latter-day Saints}
\newcommand{\usu}{SU}
\newcommand{\acu}{PCU}

\begin{abstract}
Religion is an important part of many people's lives, with reading from religious texts being among the most common and important regular practices. Modern technology has changed the way this religious reading takes place, but the interaction of technology with religious reading has not been studied in detail. In this techno-spiritual study, we surveyed students from two universities (n = 407) to ask individuals how they use technology in their personal study of Christian religious texts and Christian religious education courses. We found that paper scriptures are still frequently used even with the prevalence of technology, especially when the reader is seeking a spiritual experience. Additionally, individuals do not strictly use one particular medium; they choose different media depending on the goals of their study session. Finally, we explore benefits and drawbacks of technology in religious education courses.

\end{abstract}



\received{20 February 2007}
\received[revised]{12 March 2009}
\received[accepted]{5 June 2009}

\maketitle

\section{Introduction}

Digital technology is an increasingly important part of religious experience \cite{bell2006no, campbell2017religious}. These practices and the digital interactions surrounding them are part of an emerging sub-field of Human-Computer Interaction called ``techno-spirituality'' \cite{buie2019letus, wolf2026spirituality}. Researchers have recently explored interactions through experiences such as online worship services \cite{wolf2022spirituality, wolf2023God}, prayer \cite{song2025walking}, online giving \cite{thompson2024technology}, support for online community \cite{claisse2023keeping}, impacts on home life \cite{wyche2009extraordinary}, and technology interventions to support health of congregants \cite{oleary2022church}. Although techno-spirituality research explores all forms of spirituality and religion, much of it has focused on Christianity, which is the world's largest religion \cite{wolf2026spirituality}. In particular, a thread of research has focused on the experience of Black American Christians with religious technology~\cite{smith2026understanding, thompson2024technology, sackitey2023everyone}.

However, there are still many gaps in this emerging field. Perhaps one reason is a bias exhibited by many researchers and reviewers who do not think techno-spirituality work is part of HCI, resulting in researchers feeling that they are on the margins \cite{wolf2026spirituality}. Others cite the lack of clarity in techno-spirituality studies regarding terminology and methods~\cite{buie2019letus}. As a result, the impact of technology on many key religious practices remains unexplored. One of these areas is the reading of religious scriptures through digitally mediated means \cite{kim2022socialspiritual}. Existing research on reading has focused on the effects of digital reading materials on comprehension, distraction, and other similar topics ~\cite{goodwin2020digital,schwabe2022no,singer2017reading,LI2024100142}. The study of religious texts, however, generally has a different goal: namely, to achieve a religious experience. 

In this paper, we report on the results of a survey on the digital religious experiences of studying scripture at two universities: a mid-sized state university (SU) and a private Christian university (PCU). Our results indicate the persistence of paper scriptures, even when digital technology can be faster and provide more robust study support. However, when using digital technology, users choose different media, such as phones or laptops, for different goals. These include personal study, group study, and lesson preparation.

Our research questions are the following:
\begin{itemize}
    \item \textbf{RQ1:} How do people's spiritual experiences differ when they study their scriptures using different media?  (between paper, mobile devices, and computers)
    \item \textbf{RQ2:} How do people select different reading media  (paper/mobile device/computer) based on the specific task they are trying to accomplish?
    \item \textbf{RQ3:} What are the particular benefits people see from each of the reading media? (paper/phone/computer)
    \item \textbf{RQ4:} How do technological restrictions affect religious education courses?
\end{itemize}

We make several important contributions through this work. First, we present our study examining the impact of technology on religious experience when reading scripture. Second, our work presents techno-spiritual results on both a previously reported demographic (Protestant Christians) and a previously unreported demographic (The Church of Jesus Christ of Latter-day Saints, colloquially known as ``Mormons''). Finally, our work expands the sub-field of techno-spirituality by providing insight into when and why users choose to use technology for the religious experience of studying scripture, which is a core part of the religious experience for over a billion people.


\section{Related Work}

\subsection{The Importance of Scripture in Christian Worship}
Reading of scripture, both in and out of worship service, is considered an important metric of activity in Christian faiths, as evidenced by its inclusion as one of the primary measures of religious activity in the Pew Religious Landscape Survey~\cite{smith2025decline}. Despite recent declines, still 33\% of Christians read scriptures outside of worship on a weekly basis, with Latter-Day Saints at 57\% and Evangelical Protestantism at 51\%. A 2016 survey reports slightly different numbers for members of the Church of Jesus Christ of Latter-Day Saints~\cite{riess2019next}, with reading daily as 40\%/34.5\%/40\% for Boomer-Silent/GenX/Millennial, reading at least weekly as 26.5\%/33\%/30\% for Boomer-Silent/GenX/Millennial~\cite{riess2019next}. Just like the Pew survey, it includes the reading of religious scripture as a key metric of activity, once again showing the importance of scripture study in Christian faiths.

There are a multitude of reasons as to why Christians choose to read scriptures. Some research shows that individuals tend to read scriptures to help them with challenges in their lives. As an example, DeAngelis et al. found evidence that those of a lower socioeconomic status were more likely to read scriptures for insights on wealth, and those struggling with health were more likely to read scriptures for insights regarding health~\cite{deangelis2019scriptural}. Hamilton et al. were similarly able to show in their research that individuals found comfort for stress in different Biblical passages depending on the stressor~\cite{hamilton2013reading}. Krause and Pargament also found that regular Bible study helped individuals moderate the negative association between stress and hope~\cite{krause2018bible}. Rackley found that youths tended to study scriptures to learn about their religious tradition, to apply religious lessons to their lives, to connect to God, to find comfort, and to find strength to endure difficult times~\citep{rackley2016religious}.







\subsection{Religion and Technology}

\subsubsection{Techno-Spirituality as an HCI Research Area}
Bell's foundational ethnographic work coined the term ``techno-spiritual practices'' to describe the repurposing of information and communication technologies for religious ends, from blessed mobile phones to online genealogical services, arguing that such practices reveal alternate paradigms for ubiquitous computing design~\cite{bell2006no}. Around the same time, Wyche and Grinter interviewed American Protestant Christians about how faith shaped domestic technology use, coining the term ``extraordinary computing'' to describe systems that honor the special value households accord to religious objects and routines~\cite{wyche2009extraordinary}. Campbell, coming from a communications background, situates this body of work within Digital Religion Studies, describing how digital culture increasingly blurs the boundary between people's online and offline religious lives~\cite{campbell2017religious,campbell2022digital}. Despite this foundation, religion and spirituality remain marginal topics within HCI: Buie finds that papers on spiritual and transcendent experience frequently omit any operational definition of what they study~\cite{buie2019letus}, and Wolf et al.'s review of 206 ACM and IEEE publications and a survey of 19 researchers finds that despite growing interest, marginalization persists and the ACM and IEEE literatures remain largely disconnected from one another~\cite{wolf2026spirituality}.

\subsubsection{Technology Adoption in Religious Institutions and Communities}
Many Christian denominations, including those studied in this work, have long used emerging publishing and dissemination technologies to spread their message~\cite{peters2015mormonism}, though adoption is often uneven and shaped by each community's needs. Access is one recurring limit: Thompson et al. found that unequal access to devices and internet constrained how much technology could help Black church leaders reach their congregations~\cite{thompson2024technology}, while Sackitey et al. found that online interactions extended members' existing offline social support networks, including practices like scripture reading, but were hard to sustain without dedicated technical support~\cite{sackitey2023everyone}. Attitudes toward technology can be similarly mixed: Grinter et al. found that Protestant ministers valued technology for engagement but faced pushback from those who saw it as prioritizing entertainment over spiritual content~\cite{grinter2011technology}, and Claisse and Durrant found that video-conferencing let a Buddhist community keep meeting during COVID-19 but introduced distractions and could not reproduce the embodied experience of practicing together in person~\cite{claisse2023keeping}. On the other hand, Rifat et al. show how digital Islamic sermons can spread religious thought more widely~\cite{rifat2022situating}, and O'Leary et al. found that the most-used features of a Black church health app were its scripture-focused ones, with Bible study offering comfort during the pandemic~\cite{oleary2022church}.

\subsubsection{Individual and Demographic Patterns in Techno-Spiritual Practice}
Beyond institutional adoption, a complementary line of work examines how individuals' own techno-spiritual habits, including their reading of scripture specifically, vary by demographic group and by the kind of religious purpose being served. Muralidharan et al. found that U.S. Christians' smartphone use tracked with religiosity and supported existing practices like Bible reading and prayer, while for the Israeli Jewish participants they surveyed, who valued spirituality over religiosity, using a smartphone for a spiritual process instead increased subjective well-being relative to general use~\cite{muralidharan2023digitalization}. At the level of individual practice, Wyche and Grinter's home tours found that participants annotated personal paper Bibles kept in private spaces for quiet daily reading, while weekly Bible study was instead a group activity coordinated over email~\cite{wyche2009extraordinary}, and Song's autoethnography of Christian prayer paired a physical Bible and church materials with a YouVersion reading plan, noting that pastors still encourage carrying a physical Bible even as Bible apps have become widely accepted~\cite{song2025walking}.

A highly relevant and concurrent work to our own is
Smith et al.'s mixed-methods study of Black Christian young adults. Presented at CHI '26, they found that 81\% of respondents used a Bible app (most often YouVersion) and 58\% used one weekly, that highlighting features made the app feel more like a physical Bible, and that a gamified reading ``streak'' made some participants anxious or feel it was an inauthentic way to track scripture reading~\cite{smith2026understanding}. Our work extends theirs along several axes: rather than a primarily Black population, we survey and compare two university populations spanning a wide range of Christian traditions; rather than centering on the binary choice between a physical Bible and a Bible app, we measure medium choice, including laptops, across a specific set of purposes such as personal study, group study, and lesson preparation, and quantify how that choice shifts with purpose and session length. We also extend the inquiry beyond individual devotional practice into the classroom, examining how restricting technology in religious education courses changes the study experience. Together, this body of work suggests that religious individuals and communities continue to value non-digital elements of practice for particular purposes, a pattern we investigate directly, and with finer purpose- and population-level granularity, for the practice of scripture reading in this paper.

\subsection{Reading in Digital vs. Paper media}
The effects of reading with digital media opposed to paper have been frequently studied, with the results occasionally being contradictory. In a literature review, Singer and Alexander (2017) acknowledge a number of confounding factors in measuring comprehension which make a consensus difficult to come to~\cite{singer2017reading}. Different studies contain different definitions of comprehension, different lengths of passages, differences in the questions asked, different measures of comprehension, and even different age groups being surveyed. When readers of age 5 to 6 read simple texts, there was no difference in comprehension depending on medium, but when high-school-aged readers attempted to read more complex passages the medium began to display an effect on comprehension, with technology leading to a quicker but shallower reading of the text. Schwabe et al. (2022) assert that, in regards to narrative texts, there are no negative effects of reading via screen, and that there may perhaps even be some advantages~\cite{schwabe2022no}. They themselves point out how this is in opposition to what other studies have claimed, and point towards other meta-analyses not differentiating between text genres as the explanation for differing answers.  

Prior research also shows digital and paper media being engaged with in different ways. Goodwin et al. (2020) found in the course of their research that students were more than twice as likely to highlight and more than five times as likely to annotate when reading with paper compared to reading digitally~\cite{goodwin2020digital}. In a meta-analysis, Li and Yan (2024) found that paper reading was more suited to literature and longer passages, while digital reading thrived in situations where readers used reading strategies such as note taking and finding word definitions, along with taking advantage of interactive functions in the digital materials~\cite{LI2024100142}.

Simply put, the existing body of research shows that the differences between the use and effects of digital and paper reading materials are complex and ripe for exploration, even before adding a religious component to the study.



\section{Methods}
Due to the highly personal nature of the study of religious texts, we needed to collect data relating to people's personal experiences for this study.  We designed a survey to that effect, as it could capture the everyday experiences of individuals and their religious habits without our interference. The full text of the survey is provided in Appendix \ref{sec:survey}.

\subsection{Survey Design}
\label{sec:survey-design}
\subsubsection{Media Categories}
For the survey, we decided to divide the media into three categories: paper scriptures, phones and tablets, and laptops and desktop computers.  A wealth of research has shown that readers engage with paper materials differently than digital ones, and other studies have shown that users prefer phones and desktop computers for different tasks~\cite{Adepu2016}.  While we could have potentially further divided the categories to separate phones and tablets or desktop and laptop computers, we felt that the increased length of the survey would have deterred more participants without providing significantly different results.

\subsubsection{Personal Scripture Study and Its Purposes}
The first questions of the survey attempted to understand participants' behavior in regards to personal study of religious texts.  We started by asking for what purposes they read scriptures regularly, details on how regularly they read scriptures with each medium, and for how long, and how often they had spiritual experiences while using each medium for study.  These baseline questions were used to attempt to understand what study of religious texts looks like on an individual-to-individual basis. When asking questions about spiritual experiences, we intentionally left the definition of a spiritual experience vague. As mentioned in The Varieties of Religious Experience by William James, there are different expectations as to what a religious (or spiritual) experience entails~\cite{james2003varieties}. Since different religions, and even individuals within the same religion, have different ideas about what a spiritual experience is, we believed over-defining would exclude some types of experiences and would make the survey less generalizable. 



The purposes that could be selected included preparing for a lesson or sermon, personal study, family study, seeking an answer to a question, seeking revelation, to feel a spiritual connection, for self improvement, or some other reason that the participant could specify.  These purposes were largely determined by the researchers, though there is a large overlap with the purposes found by Rackley~\cite{rackley2016religious}.


One source that we turned to in order to find reasons why people studied scriptures was the scriptures themselves, as we assumed that those who read scriptures habitually would agree with the reasoning in them to some degree.  For example, in the Bible, 2 Timothy 3:16-17 it says ``All scripture is given by inspiration of God, and is profitable for doctrine, for reproof, for correction, for instruction in righteousness: That the man of God may be perfect, thoroughly furnished unto all good works.''  From this we extrapolated that self improvement is a reason that people study scripture. 
Additionally, local religious leaders suggests personal and family study of the scriptures, which gave us additional purposes for our study. Lastly, \churchname{}, which the overwhelming majority of \usu{} respondents belong to, and Churches of Christ, which is affiliated with \acu{}, are both volunteer run organizations. As such, members of these churches are occasionally called upon to prepare lessons or sermons for other members of their congregations, so we felt it necessary to include this as a potential purpose of scripture study as well.


\subsubsection{Media Preference and Distraction}
We hypothesized that, given the advantages and disadvantages of each medium, individuals would use different media for different purposes, and we wished to capture this reasoning in our survey.  For each purpose they chose for studying scriptures, they were asked follow-up questions about how often they used each medium for that purpose via a 5 point Likert scale, and then a short answer question about why they prefer to use the medium they do for that purpose.
According to prior work, digital devices can have immense potential to distract their users to a degree that potentially negates the advantages that digital reading offers~\cite{liu2022reading}. In order to see what effect digital distraction has on religious reading, we included a question about how frequently the participants experienced distraction with each of the listed media.

\subsubsection{Technology and Religious Education Classes}

Another context in which the personal study of religion occurs, apart from the study of scripture, is in religious education classes. At \acu{} in particular, undergraduate religion courses including the study of the Bible are required. 

We had observed that some of these classes restricted the use of technology, which prompted us to develop questions to understand the effect that technology, or its absence, has on these classes.  In order to understand this, we asked those who had experience in such classes where technology was restricted, about what kind of class it was, what restrictions were in place, and how those restrictions impacted their learning.

In the context of religious education classes, technology could have additional uses, such as for reading sources not present in paper scriptures, but it could also have additional drawbacks, such as a student's technological misuse distracting neighboring students.  Keeping these in mind, we included questions about how often the participant's learning experience in a religious education class was disrupted by their or someone else's use of technology, along with what advantages and disadvantages they saw with the use of technology in such classes. 

\subsubsection{Demographics}
We included questions about demographics for two main reasons.  The first was to understand the population we were surveying and how the results are affected by its distribution.  The second reason was to monitor recruitment so that we could ensure equal responses across majors.  More details of the recruitment process are given in Section~\ref{sec:recruitment}.

\subsection{Recruitment}
\label{sec:recruitment}

Participants were offered a \$15 Amazon gift card upon completion of the survey. To recruit participants for the survey, we sent emails primarily to professors teaching general education classes and asked them to send a Canvas message to their students advertising the survey.  With one of the classes, a researcher visited the class in person to explain the survey and ask the students to participate.  We mostly recruited from general education classes and introductory courses to try and obtain a wide selection of majors.  
We also recruited from religion courses, as religion course instructors were likely to agree to send out the survey, and the courses also have a diversity of majors.

Emails were sent in waves, where a few professors were messaged at a time, then after results had come in, the demographics were analyzed. From \usu{}, we initially recruited from religion and history courses, introductory science courses, and computer science courses. We noted that certain majors such as computer science were overrepresented, with others being underrepresented, and so with the next wave of messages, we tried to focus on professors teaching courses with underrepresented majors. In turn, we additionally recruited from introductory art, business, psychology, and human development courses. This process of emailing professors and analyzing demographics was repeated until we had reached the desired 300 survey participants from \usu. A similar process took place once the survey was extended to \acu{}, where we recruited from Bible, business, and computer science courses and recruited an additional 107 participants.
The \usu{} responses were collected between December 2024 and February 2025. The \acu{} responses were collected between November 2025 and February 2026, using the same survey instrument. All data collection was therefore complete before the concurrent work of \citet{smith2026understanding} was published.

Because this work concerns the study of Christian scripture, and because too few participants reported any other tradition to support conclusions about scripture study within it, we excluded from all subsequent analysis those participants who did not report affiliation with a Christian denomination. This removed 12 participants at \usu{} and 3 at \acu{}. Most of these participants reported no religious affiliation at all, describing themselves as atheist, agnostic, or unaffiliated; the remainder consisted of two participants who identified as Muslim, one who reported a Bah\'a'\'i affiliation, and a small number whose free-text answers did not indicate a Christian affiliation. Participants who selected ``Prefer not to say'' were retained, as were those whose free-text answers named a Christian denomination not listed among the response options. Every result we report is therefore based on the remaining 288 participants at \usu{} and 104 at \acu{}, and these are the counts given in the figures and tables throughout the paper.

\begin{figure}
    \centering
    \includegraphics[page=1, width=1\linewidth]{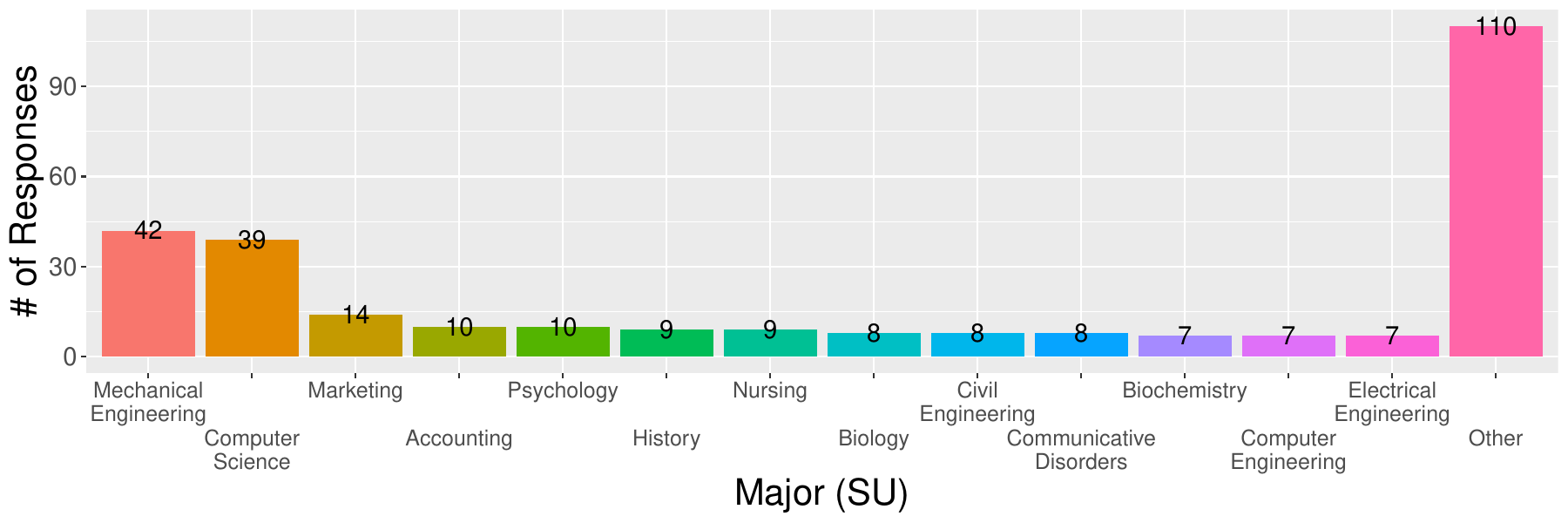}
    \includegraphics[page=2, width=1\linewidth]{figures/demographics/majors.pdf}
    \caption{Distribution of academic majors among survey participants at \usu{} (top) and \acu{} (bottom). Note the differing vertical scales. Majors with few responses are grouped together as ``Other''.}
    \label{fig:major-spread}
\end{figure}


\begin{figure}
    \centering
    \includegraphics[page=1, width=0.49\linewidth]{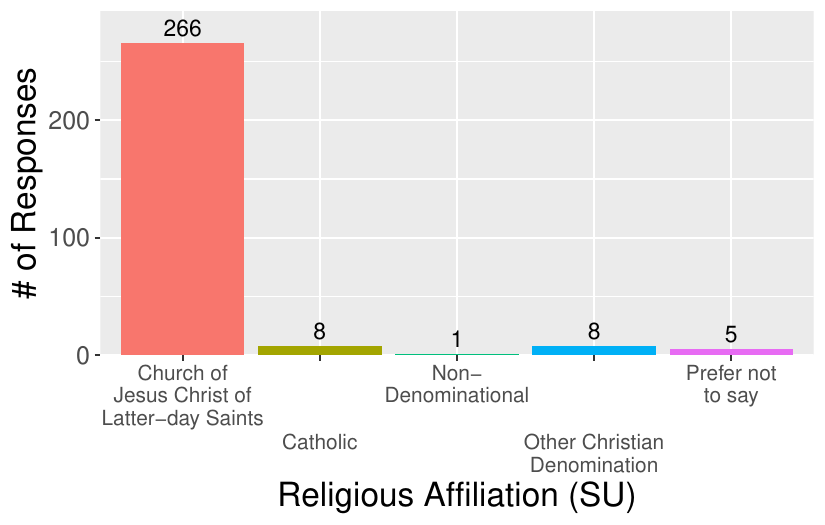}\hfill
    \includegraphics[page=2, width=0.49\linewidth]{figures/demographics/religion.pdf}
    \caption{Religious affiliation of survey participants at \usu{} ($n=288$, left) and \acu{} ($n=104$, right). Note the differing vertical scales. The \acu{} results are far more denominationally diverse than those from \usu{}, where nearly all respondents identified with \churchname{}.}
    \label{fig:religion-spread}
\end{figure}

\subsubsection{Demographics of Respondents}
A spread of the major demographics surveyed is shown in Figure~\ref{fig:major-spread}. The distribution of denominational affiliations for each university is shown in Figure~\ref{fig:religion-spread}.
Participants also reported their race/ethnicity, and were allowed to select more than one option, so the percentages reported below do not sum to 100\%. At \usu{} ($n=288$), the vast majority of participants identified as White/Caucasian (279, 96.9\%), with smaller numbers identifying as Hispanic (12, 4.2\%), Asian/Pacific Islander (1, 0.3\%), Black or African American (1, 0.3\%), Native American (1, 0.3\%), or preferring not to say (2, 0.7\%). At \acu{} ($n=104$), participants were more racially diverse though still predominately White: 73 (70.2\%) identified as White/Caucasian, 22 (21.2\%) as Hispanic, 8 (7.7\%) as Black or African American, 6 (5.8\%) as Asian/Pacific Islander, 4 (3.8\%) as Native American, 2 (1.9\%) as another race/ethnicity not listed, and 3 (2.9\%) preferred not to say.

Participants at both universities were fairly evenly split by gender and were overwhelmingly of traditional college age. At \usu{} ($n=288$), 168 (58.3\%) described themselves as men and 120 (41.7\%) as women, with no participants selecting non-binary, another gender not listed, or preferring not to say. At \acu{} ($n=104$), 57 (54.8\%) described themselves as men, 44 (42.3\%) as women, and 3 (2.9\%) preferred not to say. As for age, 224 (77.8\%) of \usu{} participants and 97 (93.3\%) of \acu{} participants were between 18 and 22 years old, with a further 50 (17.4\%) and 4 (3.8\%), respectively, between 23 and 25. Only 17 participants in total (14 at \usu{} and 3 at \acu{}) were older than 25.


\subsection{Quantitative Methods}

All of the frequency questions used the same five point Likert scale, and before any statistical tests were run we converted these responses to numerical data by coding the response options in order as `Almost Never' $= 1$, `Rarely' $= 2$, `Sometimes' $= 3$, `Often' $= 4$, and `Very Often' $= 5$. Responses left blank were excluded from the tests involving that item. Because the response options describe frequency and the same scale was used for every item, we treated the coded values as interval data so that means ANOVA (and the effect sizes described below) could be computed; the underlying distributions of the ordinal responses are shown in Figures~\ref{fig:likert-all-tasks} and~\ref{fig:likert-distr-spirit}.

With our statistical analysis, we wanted to know whether there were significant differences between the universities surveyed, between the media used for study, and between study purposes. For differences between media used and study purposes we used analysis of variance (ANOVA). For the differences between universities, a $t-$test was sufficient, as there were only the two universities surveyed; because the two samples are independent and of unequal size, we used Welch's $t-$test, which does not assume equal variances. Additionally, we wished to measure effect size so that we could determine which results were of the most practical significance. For the ANOVA tests of differences between media and between study purposes, we use $\eta^2$ as the effect size. For testing differences between universities, the raw mean difference was used to give an idea of the actual difference in values, and Hedges' $g$ was used as it is advantageous when the populations are of different sizes, as ours were. We also made various plots to allow for visual comparison of differences across variables.



\subsection{Qualitative Methods}
\label{sec:qual-methods}

We analyzed the free-response items in the survey using grounded thematic analysis~\cite{saldana2009coding}, in which the codes are built up inductively from participants' own language rather than being fixed in advance from a theoretical framework. This was a deliberate choice given our research questions: because little prior work describes how people choose between paper and digital scriptures for particular kinds of religious reading, we did not want to constrain what participants could tell us to a set of categories we had anticipated.

All coding was done by two coders, one based at each university, so that perspectives from both institutions were represented in the development of the codebook. Rather than analyzing each survey item in isolation, we grouped items of a similar type and analyzed each group together, since the same underlying reasons recurred across items within a group. The first group consisted of the free-response items asking why participants preferred a given medium when reading for each of the study purposes they had selected. The second group consisted of the items about religious education classes: the advantages and disadvantages of technology in such classes, which restrictions on technology participants had experienced, and how those restrictions affected their learning.

For each group, we followed the same three-stage process. First, both coders independently open-coded an initial subset of responses, writing descriptive codes directly from the text, and then met to compare their codes, merge overlapping ones, discard codes that appeared only once or twice, and agree on a shared codebook in which every code had a written definition and representative examples drawn from the responses themselves. The codebook for the first group was developed from the first 50 responses, and the codebook for the second group from the first 25; in both cases the coders had reached saturation, i.e. were no longer encountering new recurring themes by the end of that subset. Second, the two coders independently applied the resulting codebook to a further round of previously uncoded responses---30 responses for the first group and 20 for the second---and we measured agreement between them using Cohen's $\kappa$~\cite{cohen1960coefficient}. Agreement was $\kappa = 0.88$ for the first group of questions and $\kappa = 0.83$ for the second, both of which fall in the ``almost perfect'' range on the conventional interpretation of $\kappa$~\cite{landis1977measurement}. 
Third, having established that agreement, the coders divided the remaining responses between them and coded them separately, discussing any responses that were difficult to categorize. Codes were not mutually exclusive: a single response could be assigned several codes if it raised several distinct reasons, or none if it was blank or uninformative.
The codebooks for both groups are given in Tables~\ref{tab:codebook-1} and~\ref{tab:codebook-2}, with example responses for each code.

\subsection{Limitations}
\label{sec:limitations}

\subsubsection{Self-report} Our measures are participants' accounts of their practice rather than observations of it, and reports of frequency and duration in particular require aggregating many past sessions from memory. Our session and weekly study lengths are therefore best read as relative comparisons between media rather than accurate estimates of time spent, and asking which medium serves each purpose may elicit a tidier account than the practice behind it; telemetry or a diary study would be needed to confirm these patterns. Religious practice also adds a social desirability risk, as framing a survey around scripture study may have prompted some participants to present themselves as more devout than they are. 

\subsubsection{Population} Participation was voluntary and incentivized, so students who already study scripture, or who hold views about technology's effect on it, had more reason to respond---a bias toward the religiously engaged that recruiting in religion courses likely amplified. Our participants were also overwhelmingly of traditional college age, predominantly White, and drawn from a single region of the United States, so we cannot speak to older adults, children learning to read scripture, or non-Western contexts, and our religious education findings are specific to the two institutional arrangements described in Section~\ref{sec:rq4}. The sample is also Christian by construction (Section~\ref{sec:recruitment}), which excludes other traditions and the religiously unaffiliated. 

\subsubsection{Analysis} We treat university, medium, and study purpose as our variables of interest and did not break results down by age, gender, race, or major, so variation along those lines is invisible in our results. Our two universities also differ in institution, denominational makeup, region, and religious education structure all at once, so we can show that the two populations behave differently but not isolate which factor is responsible. Finally, grouping phones with tablets and laptops with desktops, done to keep the survey short, hides any differences within those pairings. We suggest analysis along these lines would be fruitful future work.

\section{Results}

\subsection{Quantitative Results}
\begin{figure}
    \centering
    \includegraphics[width=1.1\linewidth]{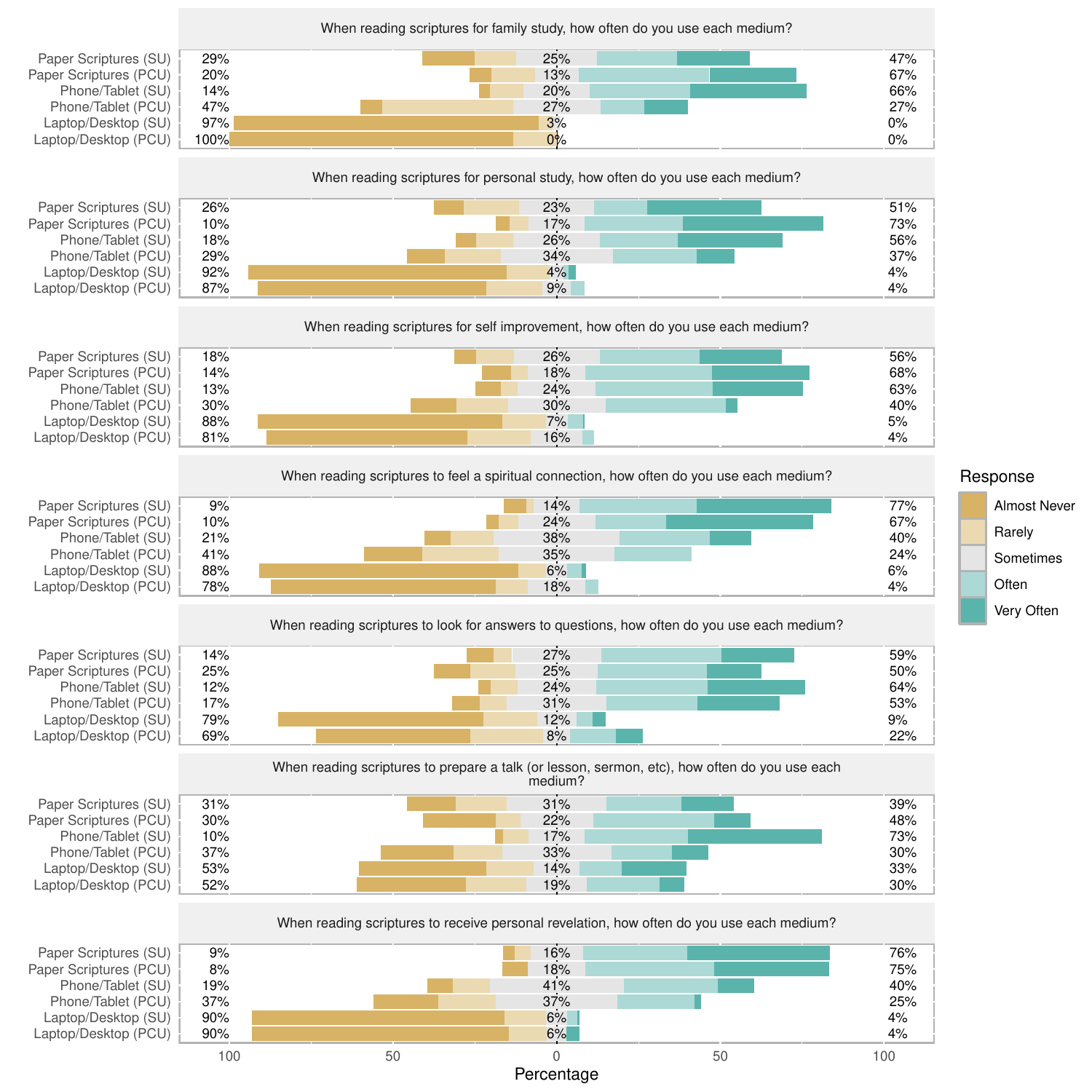}
    \caption{How often scriptures are read with each medium for a given task, for each university.}
    \label{fig:likert-all-tasks}
\end{figure}

\begin{figure}
    \centering
    \includegraphics[width=1.1\linewidth]{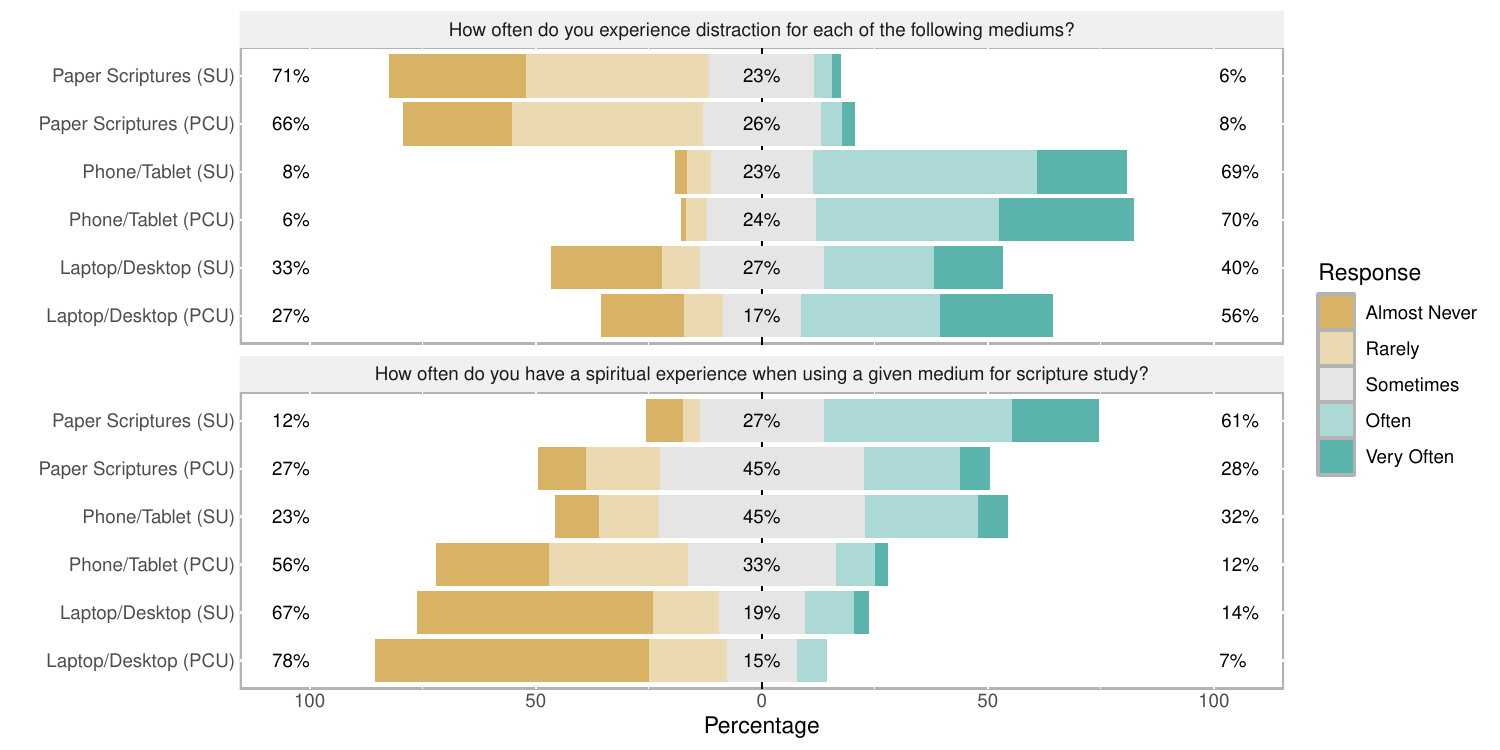}
    \caption{How often participants experience distraction and have spiritual experiences with each medium, for each university.}
    \label{fig:likert-distr-spirit}
\end{figure}

The distributions of responses to quantitative questions are shown in Figures~\ref{fig:likert-all-tasks} and \ref{fig:likert-distr-spirit}. The former shows how often participants read scriptures with each medium for each purpose, and the latter shows how often they experienced distraction or a spiritual experience with each medium. The distributions are broken down by university.
As shown in Tables \ref{tab:p-methods}, \ref{tab:p-univ}, and \ref{tab:p-questions}, many of the results were statistically significant. This is unsurprising, due to the large study population surveyed and the number of questions asked. As a result, we will focus our discussion on the results with large effect sizes, which are of more practical significance.

\begin{table}[h!]
\begin{center}
\begin{tabular}{ |c|c|c|c|c| } 
 \hline
    & \multicolumn{2}{|c|}{\usu{}}&\multicolumn{2}{|c|}{\acu{}}\\
 \hline
& $p-$value& $\eta^2$& $p-$value& $\eta^2$\\
\specialrule{1.75pt}{0pt}{0pt}
Spiritual Experience Freq & \textbf{$<$ 0.001}& 0.268 & \textbf{$<$ 0.001}& 0.213\\ \hline
Prepare for Talk & \textbf{$<$ 0.001}& 0.163 & 0.320 & 0.029\\ \hline
Personal Study & \textbf{$<$ 0.001}& 0.453 & \textbf{$<$ 0.001}& \underline{0.506}\\ \hline
Family Study & \textbf{$<$ 0.001}& \underline{0.565} & \textbf{$<$ 0.001}& \underline{0.540}\\ \hline
Find Answers & \textbf{$<$ 0.001}& 0.417 & \textbf{$<$ 0.001}& 0.190\\ \hline
Seek Revelation & \textbf{$<$ 0.001}& \underline{0.573} & \textbf{$<$ 0.001}& \underline{0.500}\\ \hline
Feel Connection & \textbf{$<$ 0.001}& \underline{0.527} & \textbf{$<$ 0.001}& 0.480\\ \hline
Self improvement & \textbf{$<$ 0.001}& 0.487 & \textbf{$<$ 0.001}& 0.409\\ \hline
Distractions & \textbf{$<$ 0.001}& 0.289 & \textbf{$<$ 0.001}& 0.294\\
 \hline
\end{tabular}
\caption{Differences in how often the three media are used, tested separately for each university and item (bold: $p < 0.05$; underline: $\eta^2 > 0.5$). What participants are reading for strongly shapes which medium they use. Figure~\ref{fig:likert-all-tasks} shows the distributions behind these numbers.}
\label{tab:p-methods}
\end{center}
\end{table}

\begin{table}[h!]
\begin{center}
\resizebox{1\linewidth}{!}{%
\begin{tabular}{ |c|c|c|c|c|c|c|c|c|c| } 
 \hline
    & \multicolumn{3}{|c|}{Paper}&\multicolumn{3}{|c|}{Phone/Tablet}&\multicolumn{3}{|c|}{Laptop/Desktop} \\
 \hline
& $p-$value& Mean&Hedges'& $p-$value& Mean&Hedges'& $p-$value& Mean&Hedges'\\
&& Diff&$g$&& Diff&$g$&& Diff&$g$\\
\specialrule{1.75pt}{0pt}{0pt}
Spiritual Experience Freq & \textbf{$<$ 0.001}& 0.630 & 0.585 & \textbf{$<$ 0.001}& \underline{0.719} & \underline{0.701} & \textbf{ 0.010 }& 0.307 & 0.266\\ \hline
Prepare for Talk & 0.945 & 0.019 & 0.015 & \textbf{$<$ 0.001}& \underline{1.190} & \underline{1.084} & 0.775 & 0.083 & 0.053\\ \hline
Personal Study & \textbf{ 0.001 }& -0.509 & -0.388 & \textbf{$<$ 0.001}& 0.552 & 0.456 & 0.281 & -0.120 & -0.146\\ \hline
Family Study & 0.226 & -0.429 & -0.316 & \textbf{ 0.007 }& \underline{0.981} & \underline{0.864} & 0.684 & -0.040 & -0.108\\ \hline
Find Answers & 0.215 & 0.283 & 0.243 & 0.257 & 0.251 & 0.226 & 0.087 & -0.430 & -0.370\\ \hline
Seek Revelation & 0.467 & 0.128 & 0.119 & \textbf{ 0.003 }& 0.533 & 0.499 & 0.888 & -0.020 & -0.024\\ \hline
Feel Connection & 0.812 & 0.042 & 0.037 & \textbf{$<$ 0.001}& 0.588 & 0.543 & 0.248 & -0.166 & -0.183\\ \hline
Self improvement & 0.284 & -0.192 & -0.162 & \textbf{$<$ 0.001}& \underline{0.705} & \underline{0.614} & 0.159 & -0.185 & -0.216\\ \hline
Distractions & 0.226 & -0.132 & -0.141 & 0.156 & -0.148 & -0.162 & \textbf{ 0.019 }& -0.384 & -0.274\\
 \hline
\end{tabular}
}
\caption{Differences between the universities in how often each medium is used. Mean differences are \usu{} minus \acu{} on the five-point scale, so positive values mean \usu{} reported the higher rate (bold: $p < 0.05$; underline: the four largest effects, $|g| \ge 0.6$, all for phones and tablets). The universities differ chiefly in digital use, not paper use. See Figure~\ref{fig:likert-all-tasks} for the distributions behind the study-purpose rows and Figure~\ref{fig:likert-distr-spirit} for the spiritual experience and distraction rows.}
\label{tab:p-univ}
\end{center}
\end{table}

\begin{table}[h!]
\begin{center}
\begin{tabular}{ |c|c|c|c|c| } 
 \hline
    & \multicolumn{2}{|c|}{\usu{}}&\multicolumn{2}{|c|}{\acu{}}\\
 \hline
& $p-$value& $\eta^2$& $p-$value& $\eta^2$\\
\specialrule{1.75pt}{0pt}{0pt}
Paper Scriptures & \textbf{$<$ 0.001}& 0.069 & \textbf{ 0.002 }& 0.065\\ \hline
Phone/Tablet & \textbf{$<$ 0.001}& 0.056 & \textbf{ 0.012 }& 0.053\\ \hline
Laptop/Desktop & \textbf{$<$ 0.001}& \underline{0.169} & \textbf{$<$ 0.001}& \underline{0.118}\\
 \hline
\end{tabular}
\caption{Differences across the seven study purposes (preparing for a talk through self improvement), tested separately for each medium and university (bold: $p < 0.05$; underline: $\eta^2 > 0.1$). Purpose shapes how often a given medium is reached for at both universities, but far less strongly than the choice of medium itself does in Table~\ref{tab:p-methods}.}
\label{tab:p-questions}
\end{center}
\end{table}

\subsubsection{Differences Between Media}
When comparing the different media used during study, nearly every test was highly statistically significant, with high effect sizes as well. In Table~\ref{tab:p-methods}, the only non-significant test is preparing for a talk at \acu{} ($p = 0.32$). This non-significant test is notable in that it means \acu{} students use the three media at comparable rates when preparing for a talk, even though they used paper scriptures at a much higher rate for every other purpose. The large effect sizes can in large part be attributed to the extremely low usage of laptops and desktops for scripture study in general, as shown in Figure~\ref{fig:likert-all-tasks}. For every purpose of scripture study, for both universities, laptops and desktops were less likely to be used than other media, and never had more `often' and `very often' results than `rarely' and `almost never' results. Between these tasks, however, there were two purposes for which participants were relatively more likely to use laptops and desktops. When reading to find answers to a question, nearly 10\% of \usu{} participants and 25\% of \acu{} participants were more likely than not to use laptops and desktops, and when reading to prepare for a lesson or sermon, that number jumped to nearly a third for both universities.

\subsubsection{Differences Between Study Purposes}
The purpose of study also shapes how often a given medium is reached for, as every test in Table~\ref{tab:p-questions} is significant, but these effects are much smaller than those for the medium itself in Table~\ref{tab:p-methods}: which medium is used is a stronger distinction than what it is used for. Purpose matters most for laptops and desktops ($\eta^2 = 0.169$ at \usu{}, $0.118$ at \acu{}), the medium that is otherwise little used and is turned to mainly for the two purposes just described. Paper and phones or tablets show smaller effects ($0.053$ to $0.069$), as both are used across every purpose.

\subsubsection{Differences Between the Universities}
In general, the usage of paper scriptures between the universities is not statistically significant. As Table~\ref{tab:p-univ} shows, paper use differs on only two of the nine items, and in opposite directions---\usu{} higher for frequency of spiritual experiences, \acu{} higher for personal study. The usage of phones and tablets for scripture study, on the other hand, shows a notable effect size, and is higher at \usu{} for seven of the nine items, most sharply for preparing for a talk (mean difference $1.19$, $|g| = 1.08$, the largest effect in our data). For the majority of tasks, \usu{} participants were more likely to use phones and tablets than paper scriptures, with the two purposes which were exceptions being feeling a spiritual connection and receiving personal revelation. For these two purposes, \usu{} participants were much more likely to use paper scriptures, and these two purposes are additionally the ones with an inherent spiritual component. \acu{} participants were much more likely to use paper scriptures for nearly every task, with the exception being looking for answers to questions. While phones and tablets were barely more common than paper scriptures for this task, it is still notable given that paper scriptures were by far favored by \acu{} participants for other tasks. The one non-significant test in Table~\ref{tab:p-methods} is similarly revealing: \acu{} participants used the three media at comparable rates when preparing for a talk, even though they used paper scriptures at a much higher rate for every other purpose.

\subsubsection{Distraction and Spiritual Experience}
One expected result, as shown in Figure~\ref{fig:likert-distr-spirit}, is that distraction is much higher with digital media than with paper scriptures, which mirrors what prior work has found for non-religious reading~\cite{liu2022reading}. It is noteworthy however that the frequency of spiritual experiences, as shown in the same Likert scale graph, appears to have an inverse relation to the frequency of distraction experienced. Additionally, \usu{} participants were more likely than not to have spiritual experiences with a phone or tablet, while \acu{} participants were much less likely to have spiritual experiences with a phone or tablet. Furthermore, \usu{} participants had a higher rate of spiritual experiences with paper scriptures than \acu{} participants did. This is likely due to a difference in understanding and expectations of spiritual experiences between the two populations.

\subsubsection{Length and Frequency of Study Sessions}
\begin{figure}
    \centering
    \includegraphics[page=1, width=0.49\linewidth]{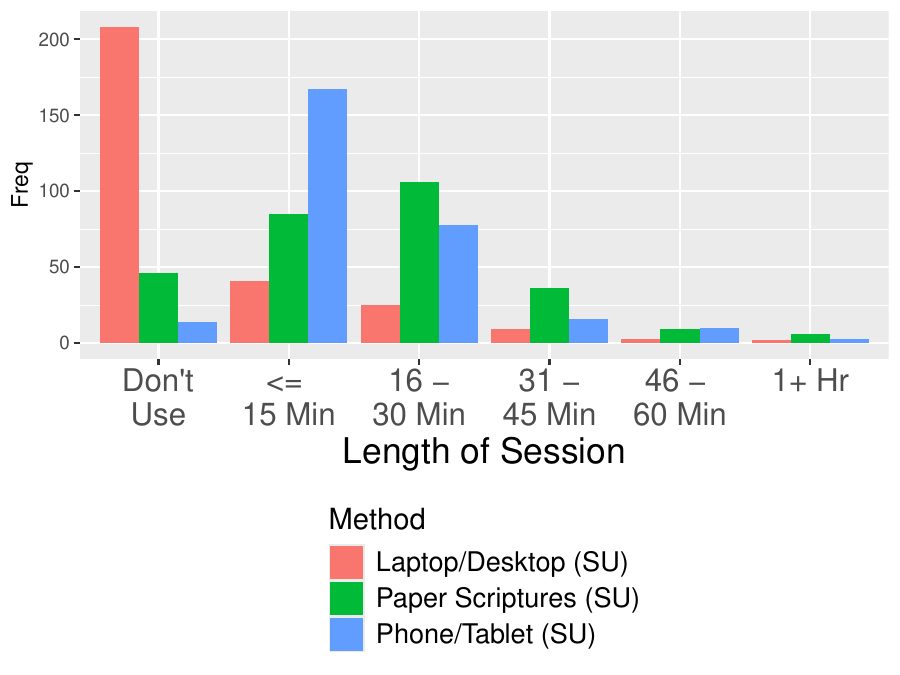}\hfill
    \includegraphics[page=2, width=0.49\linewidth]{figures/comparison/bar-plot-session-length.pdf}
    \caption{Average length of session for scripture study with each medium, for each university. Note the differing vertical scales.}
    \label{fig:bar-study-length}
\end{figure}

With Figure~\ref{fig:bar-study-length} we see that phones and tablets are most commonly used in shorter sessions, those being sessions that are 15 minutes or less. Paper scriptures are most common for sessions between 15 and 45 minutes in length. For sessions longer than 45 minutes, responses are too infrequent to draw conclusions about medium preference. Laptops and desktops were the least used medium of study for nearly every category, except for shorter sessions among \acu{} students, where it is more common than paper scriptures, but still less commonly used than phones and tablets.

\begin{figure}
    \centering
    \includegraphics[page=1, width=0.49\linewidth]{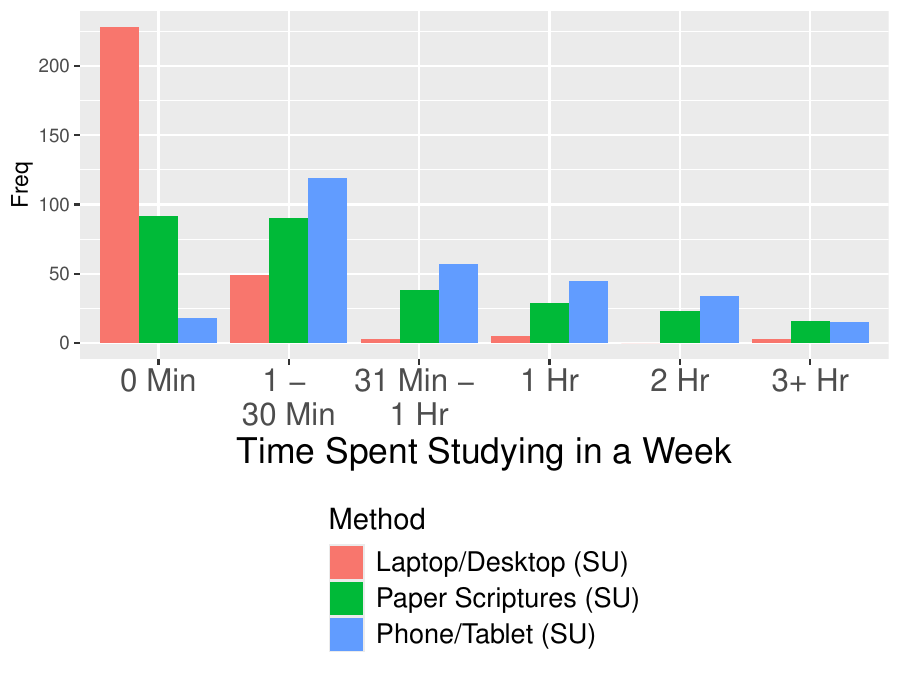}\hfill
    \includegraphics[page=2, width=0.49\linewidth]{figures/comparison/bar-plot-total-length.pdf}
    \caption{Total time spent for scripture study each week with each medium, for each university. Note the differing vertical scales.}
    \label{fig:bar-total-length}
\end{figure}

With Figure~\ref{fig:bar-total-length} we see that \usu{} participants tended to use phones and tablets more than paper scriptures for weekly study for nearly every amount of time, with paper only passing phones and tablets when the total study amount exceeded 3 hours, and even then the difference was slight. For \acu{} participants, paper was more common than phones and tablets unless the total study time was 30 minutes or less. Laptops and desktops were by far the least used, with most not using them at all.

\begin{figure}
    \centering
    \includegraphics[page=1, width=0.49\linewidth]{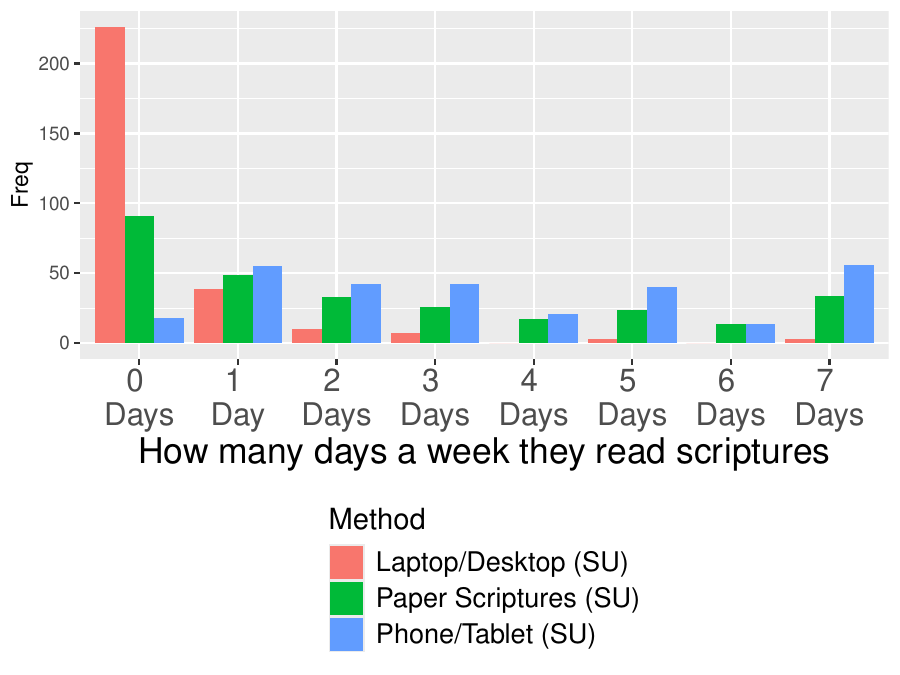}\hfill
    \includegraphics[page=2, width=0.49\linewidth]{figures/comparison/bar-plot-days-a-week.pdf}
    \caption{How many days a week scripture study occurs with each medium, for each university. Note the differing vertical scales.}
    \label{fig:bar-study-frequency}
\end{figure}

Figure~\ref{fig:bar-study-frequency} shows a similar story, with laptops and desktops rarely being used, \usu{} participants preferring phones and tablets and \acu{} participants preferring paper scriptures most of the time, with paper and phones being tied when reading 6 or 7 days a week. 

\subsubsection{Reasons for Studying Scripture}
\begin{figure}
    \centering
    \includegraphics[page=1, width=\linewidth]{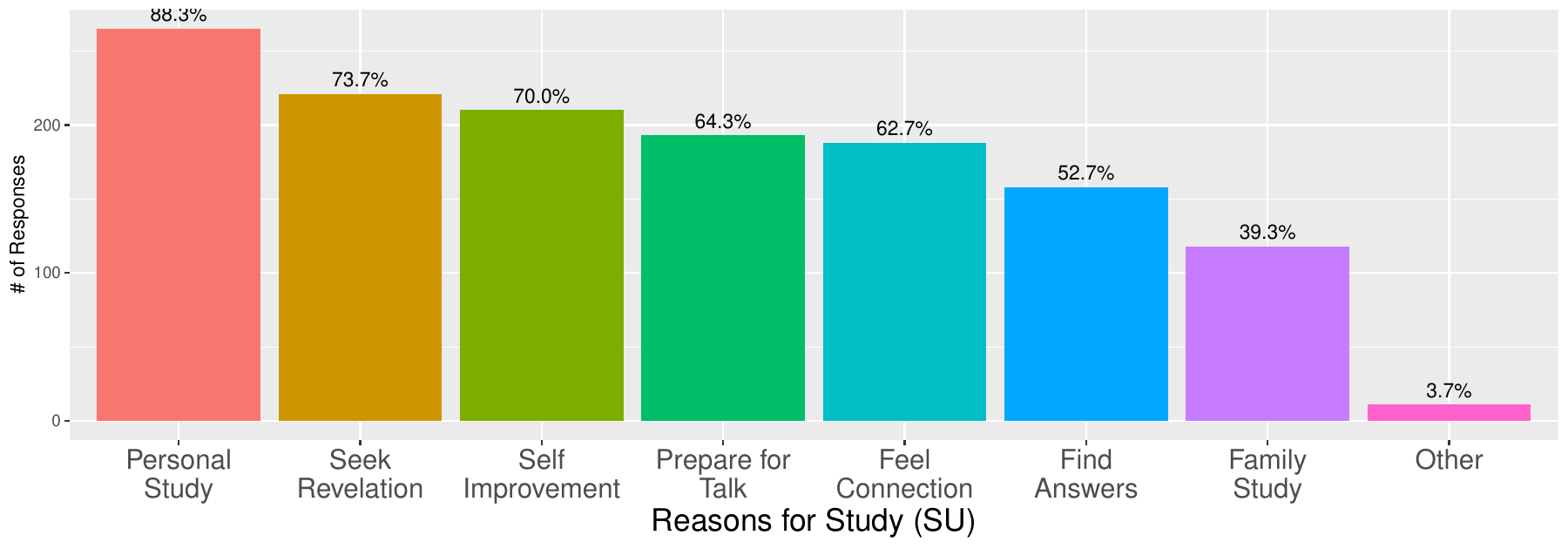}
    \includegraphics[page=2, width=\linewidth]{figures/comparison/reasons-for-study.pdf}
    \caption{The reasons given at each university for reading scriptures regularly. Participants could select more than one reason, so percentages do not sum to 100\%. Categories are ordered by their frequency at \usu{}.}
    \label{fig:bar-reasons-study}
\end{figure}

Figure~\ref{fig:bar-reasons-study} shows that \usu{} participants and \acu{} participants largely had the same distribution of purposes for studying scriptures, with personal study being the most common for \usu{} and tied as the most common with seeking revelation for \acu{}. One large difference was the number of participants who studied to prepare for talks or sermons, which is much higher in the \usu{} population than the \acu{} population. This is likely due to the structure of the churches among respondents, as \churchname{} and Churches of Christ are both volunteer-run churches and require members to occasionally give sermons, devotional thoughts, or Bible class lessons. \usu{} had nearly 90\% of its respondents identify as members of \churchname{}, while \acu{} only had 21\% of its respondents identify as members of Churches of Christ.

\subsection{Qualitative Results}
\label{sec:qual-results}

\begin{table}[h!]
\centering
\begin{tabular}{|p{2cm}|p{5.5cm}|p{6cm}|}
 \hline
    \multicolumn{3}{|c|}{Code Book 1: Codes For Specific Study Tasks} \\
 \hline
    Code&Definition&Examples from Transcripts\\
 \specialrule{1.75pt}{0pt}{0pt}
 Customary&Participant uses a given medium out of tradition or habit& ``Reading the printed Bible for me just feels like I’m reading it authentically just like many Church fathers and theologians have before me.'' (\acu{} P12)\\ \hline
 Connection&Participant feels a connection when using a given medium, typically spiritual&``I feel closer to God when I read the paper scriptures.'' (\usu{} P40)\\ \hline
 Marking/ Notes&Participant uses a given medium due to the ability to make markings or take notes on what they've read&``I love to be able to physically mark it and make notes right on the pages.'' (\usu{} P77)\\ \hline
 External \hspace{1.5cm}Resources&Participant uses a given medium due to its ability to access outside resources&``With the expanded resources from technology, I find that I spend more time in my studies learning about what I can do to make myself better.'' (\usu{} P87)\\ \hline
 Searching&Participant makes use of the search feature in a given medium&``I am able to locate specific texts online even if I don’t know exactly where they are in the scripture'' (\acu{} P31)\\ \hline
 Focus/ \hspace{1.5cm}Distractions&Participant notes that a given medium distracts them or otherwise affects their focus&``The paper scriptures allow me to ponder and focus better.'' (\usu{} P75)\\ \hline
 Convenient&Participant uses a given medium due to its convenience and accessibility&``[My] phone is always on me so it’s convenient.'' (\acu{} P35)\\ \hline
 Sensation/ \hspace{1.5cm}Tangible&Participant uses a given medium due to a preference regarding the physical sensation of it&``I have always liked reading out of paper books it helps it feel more real.'' (\usu{} P69)\\
 \hline
\end{tabular}
\caption{Code book for questions regarding the preference of media for specific scripture study tasks}
\label{tab:codebook-1}
\end{table}

\begin{table}
\centering
\begin{tabular}{ |p{2cm}|p{5.5cm}|p{6cm}|}
 \hline
\multicolumn{3}{|c|}{Code Book 2: Codes For Religion Classes} \\
 \hline
    Code&Definition&Examples from Transcripts\\
 \specialrule{1.75pt}{0pt}{0pt}
 Distractions/ \hspace{1cm}Engagement & Participant noted technology distracted or otherwise had an affect on engagement in the class&``People are easily distracted with social media and games.'' (\usu{} P28)\\ \hline
 Taking Notes&Participant mentioned use of technology affecting note taking&``Taking notes is an advantage with technologies.'' (\acu{} P96)\\ \hline
 Find Scripture Meaning&Participant noted technology was used to understand the meaning of a scripture passage&``I can quickly google a clarifying question, which helps me better understand the reading.'' (\acu{} P68)\\ \hline
 Find Scripture\hspace{1.5cm} Location&Participant noted technology was used to find the location of a scripture passage&``Faster and easier to find scriptures or other resources the teacher wants us to look up and read.'' (\usu{} P192)\\ \hline
 Convenience&Participant noted technology was used due to convenience&``It is very convenient. It is always with you.'' (\usu{} P168)\\ \hline
 External \hspace{1.5cm}Resources&Participant noted technology was used to access outside sources&``I have more access to resources that are not readily available on paper.'' (\usu{} P130)\\
 \specialrule{1.75pt}{0pt}{0pt}
 No Tech &Participant was not allowed to use technology in the class.&``No technology, only a notebook and your bible.'' (\acu{} P26)\\ \hline
 Partial Tech&Participant had some technology restrictions in the class, but had occasional use&``Phones were only out for use for scriptures, and must be put away during other times.'' (\usu{} P38)\\ \hline
 No \hspace{2.5cm}Restrictions&Participant had no notable restrictions on technology use in the class&``I have not participated in a class where technology was restricted.'' (\usu{} P29)\\
  
 \hline
\end{tabular}
\caption{Code book for questions regarding the effects of technology and restrictions of technology on religious education classes}
\label{tab:codebook-2}
\end{table}

\subsubsection{Scripture Study}

\begin{table}
\centering
\resizebox{\linewidth}{!}{%
    \begin{tabular}{|c|c|c|c|c|c|c|c|c|}
    \hline
    Code
    & \multicolumn{2}{|c|}{Customary}
    & \multicolumn{2}{|c|}{Connection}
    & \multicolumn{2}{|c|}{Marking/Notes}
    & \multicolumn{2}{|c|}{External Resources} \\
    \hline 
    Question & \usu{} & \acu{} & \usu{} & \acu{} & \usu{} & \acu{} & \usu{} & \acu{} \\
    \specialrule{1.75pt}{0pt}{0pt}
    Prepare for Talk & 4.08\% & 3.70\% & 2.55\% & 11.11\% & 28.06\% & 33.33\% & 26.53\% & 14.81\% \\ \hline
    Personal Study & 2.57\% & 8.45\% & 10.66\% & 18.31\% & 26.10\% & 14.08\% & 9.56\% & 11.27\% \\ \hline
    Family Study & 6.67\% & 13.33\% & 5.00\% & 26.67\% & 1.67\% & 0.00\% & 6.67\% & 0.00\% \\ \hline
    Find Answers & 3.14\% & 2.63\% & 11.32\% & 13.16\% & 4.40\% & 2.63\% & 20.13\% & 10.53\% \\ \hline
    Seek Revelation & 5.76\% & 9.62\% & 20.42\% & 26.92\% & 5.76\% & 1.92\% & 8.90\% & 5.77\% \\ \hline
    Feel Connection & 2.68\% & 5.88\% & 19.64\% & 27.45\% & 2.68\% & 5.88\% & 2.68\% & 3.92\% \\ \hline
    Self improvement & 0.47\% & 3.45\% & 10.85\% & 6.90\% & 7.08\% & 8.62\% & 16.51\% & 13.79\% \\ \hline
    Other & 0.00\% & 8.33\% & 0.00\% & 8.33\% & 6.25\% & 0.00\% & 0.00\% & 0.00\% \\
    \specialrule{1.75pt}{0pt}{0pt}
    Total & 11.67\% & 11.21\% & 34.00\% & 31.78\% & 43.67\% & 19.63\% & 39.33\% & 22.43\% \\
    \hline
    \hline
    Code & \multicolumn{2}{|c|}{Searching}
    & \multicolumn{2}{|c|}{Focus/Distractions}
    & \multicolumn{2}{|c|}{Convenience}
    & \multicolumn{2}{|c|}{Sensation/Tangible} \\
    \hline
    Question & \usu{} & \acu{} & \usu{} & \acu{} & \usu{} & \acu{} & \usu{} & \acu{} \\
    \specialrule{1.75pt}{0pt}{0pt}
    Prepare for Talk & 25.51\% & 22.22\% & 6.63\% & 3.70\% & 34.69\% & 29.63\% & 1.53\% & 11.11\%\\ \hline
    Personal Study & 3.68\% & 5.63\% & 26.47\% & 29.58\% & 41.18\% & 26.76\% & 9.56\% & 1.41\%\\ \hline
    Family Study & 0.00\% & 6.67\% & 18.33\% & 20.00\% & 46.67\% & 20.00\% & 1.67\% & 0.00\%\\ \hline
    Find Answers & 30.82\% & 42.11\% & 14.47\% & 2.63\% & 18.87\% & 36.84\% & 5.03\% & 2.63\%\\ \hline
    Seek Revelation & 3.66\% & 3.85\% & 38.74\% & 19.23\% & 19.37\% & 19.23\% & 7.85\% & 9.62\%\\ \hline
    Feel Connection & 0.45\% & 1.96\% & 33.48\% & 27.45\% & 20.98\% & 15.69\% & 8.93\% & 13.73\%\\ \hline
    Self improvement & 10.85\% & 8.62\% & 14.15\% & 8.62\% & 31.13\% & 25.86\% & 5.66\% & 8.62\%\\ \hline
    Other & 12.50\% & 8.33\% & 0.00\% & 0.00\% & 43.75\% & 50.00\% & 6.25\% & 0.00\%\\
    \specialrule{1.75pt}{0pt}{0pt}
    Total & 35.67\% & 22.43\% & 51.67\% & 36.45\% & 68.00\% & 52.34\% & 21.33\% & 16.82\% \\
    \hline
    \end{tabular}        
}
\caption{Frequency of specific codes for the questions regarding the preference of media for specific scripture study tasks.}
\label{tab:codes-study}
\end{table}

For reading tasks, the results of qualitative tagging as seen in Table~\ref{tab:codes-study} shows that more than half of both populations used phones or tablets for reading due in some part to the convenience and portability they offered. As one respondent noted, ``My phone is always with me, and it’s easy to access.'' (\usu{} P98) Additionally, 39\% of \usu{} respondents and 22\% of \acu{} respondents mentioned the access to external resources as a reason they used technology for reading scriptures rather than traditional paper scriptures. As an example, one respondent noted how with digital scriptures ``I can ... watch a video on scripture which helps me understand it better'' (\acu{} P75). Another advantage of digital reading materials was the search function. Searching helped participants find passages they couldn't remember the location of as well as to find passages relating to a topic they wished to study. Searching was particularly prevalent for the tasks of finding an answer to a question and preparing a lesson/sermon, as they could ``search the topic and find the scriptures related to that topic quickly'' (\usu{} P55), as one respondent put it. Another advantage noted by a few respondents was the capability to listen to the scriptures like an audio book, or as one participant put it, ``to use audio to listen rather than read.'' (\usu{} P32)

Paper scriptures were noted to have their advantages as well. 43\% of \usu{} respondents and 19\% of \acu{} respondents reported that marking scriptures and taking notes was a factor in determining which medium they used. While most who cited this as a factor said that paper was better for this particular function, a minority preferred to use online tools for marking and notes. A number of respondents also mentioned previous notes taken as a factor in choosing their medium, as being able to see notes previously taken on the page helped their study. In the words of one respondent, ``It has my markings and notations from important moments in my life and helps me remember the spiritual feelings I have felt before.'' (\usu{} P132) The feeling of connection was also a notable feature of paper scriptures, as reported by 34\% of \usu{} respondents and 32\% of \acu{} respondents. Paper scriptures were also noted to be better for avoiding distractions, with 51\% of \usu{} respondents and 36\% of \acu{} respondents reporting focus and distractions to be a factor in their choice of medium. This was particularly prevalent when being used for more personal tasks, such as receiving personal revelation and feeling a spiritual connection. 

Another interesting result came from the few participants who mentioned using multiple media simultaneously. One participant mentioned that they ``do [their] reading ... with the paper scriptures and use the laptop alongside for additional resources and music'' (\usu{} P60).

\subsubsection{Religious Education Classes}

For religious education classes, there was overwhelming consensus that technology in class causes distractions, with nearly 88\% of \usu{} respondents and 91\% of \acu{} respondents mentioning the distraction in class from technology. One respondent also noted that technology misuse in class can affect others, saying ``Those that are attempting to pay attention or add to the conversation can easily get distracted by [others using their devices]'' (\acu{} P89). Technology still had its advantages, however, as participants noted its usefulness in understanding the meaning of scriptures, finding their locations, accessing external resources, note taking, and overall convenience.

The largest effect of technology restrictions related to distractions, with 24\% of \usu{} and 47\% of \acu{} participants mentioning its effectiveness. It did have some downsides, most notably impacting the note-taking abilities of \acu{} participants. While some mentioned positive effects on their note-taking ability due to increased focus, several also voiced how removing technology inhibited their note-taking abilities. According to one participant, ``I was forced to pay attention, but I am much better at taking notes on a laptop than I am by paper. So I would often just give up trying to take notes and sit and listen.'' (\acu{} P63)  Additionally, some respondents mentioned restrictions as a demotivating factor that could lead to some not coming to class. One participant voiced the opinion that ``in general those people who come to these classes and don't pay any attention because of technology distraction would likely not come at all if there were technology restrictions.'' (\usu{} P22)

Regarding the types of restrictions experienced, \usu{} was notably more lenient, with only 33\% not allowing technology at all and 57\% having no restrictions, compared to \acu{} where 57\% did not allow technology and only 27\% had no restrictions.

\begin{table}
\centering
\resizebox{1\linewidth}{!}{%
\begin{tabular}{
|c|c|c|c|c|c|c|c|c|c|c|
}
\hline
Code
&\multicolumn{2}{|c|}{Distraction}
&\multicolumn{2}{|c|}{Taking Notes}
&\multicolumn{2}{|c|}{Find Scripture}
&\multicolumn{2}{|c|}{Find Scripture}
&\multicolumn{2}{|c|}{Convenience} \\
&\multicolumn{2}{|c|}{}
&\multicolumn{2}{|c|}{}
&\multicolumn{2}{|c|}{Meaning}
&\multicolumn{2}{|c|}{Location}
&\multicolumn{2}{|c|}{}\\
\hline
Question & \usu{} & \acu{} & \usu{} & \acu{} & \usu{} & \acu{} & \usu{} & \acu{} & \usu{} & \acu{} \\ 
\specialrule{1.75pt}{0pt}{0pt}
Tech Advantages & 0.67\% & 0.00\% & 9.67\% & 7.48\% & 3.33\% & 13.08\% & 14.00\% & 14.95\% & 52.00\% & 28.97\%\\ \hline
Tech Disadvantages & 86.67\% & 87.85\% & 0.33\% & 2.80\% & 0.00\% & 0.93\% & 1.33\% & 0.00\% & 0.00\% & 0.00\%\\ \hline
Which Restrictions & 0.00\% & 0.00\% & 0.00\% & 0.93\% & 0.00\% & 0.00\% & 0.00\% & 0.00\% & 0.00\% & 0.00\%\\ \hline
Impact of Restrictions & 23.67\% & 46.73\% & 2.00\% & 12.15\% & 0.33\% & 0.00\% & 1.33\% & 0.00\% & 1.33\% & 1.87\%\\ 
\specialrule{1.75pt}{0pt}{0pt}
Total & 87.67\% & 90.65\% & 11.33\% & 20.56\% & 3.33\% & 13.08\% & 16.33\% & 14.95\% & 52.33\% & 30.84\%\\
\hline
\hline
Code 
&\multicolumn{2}{|c|}{External Resources}
&\multicolumn{2}{|c|}{No Tech}
&\multicolumn{2}{|c|}{Partial Tech}
&\multicolumn{2}{|c|}{No Restrictions}
&\multicolumn{2}{|c|}{}\\
\hline
Question & \usu{} & \acu{} & \usu{} & \acu{} & \usu{} & \acu{} & \usu{} & \acu{} &&\\
\specialrule{1.75pt}{0pt}{0pt}
Tech Advantages & 36.00\% & 24.30\% & 0.33\% & 0.00\% & 0.00\% & 0.00\% & 0.00\% & 0.00\%&&\\ \hline
Tech Disadvantages & 0.67\% & 0.00\% & 0.00\% & 0.00\% & 0.00\% & 0.00\% & 0.00\% & 0.00\%&&\\ \hline
Restrictions Experienced & 0.00\% & 0.00\% & 32.67\% & 57.01\% & 6.33\% & 9.35\% & 57.00\% & 27.10\%&&\\ \hline
Impact of Restrictions & 2.67\% & 2.80\% & 0.00\% & 0.00\% & 0.00\% & 0.00\% & 0.00\% & 0.00\%&&\\
\specialrule{1.75pt}{0pt}{0pt}
Total & 36.33\% & 25.23\% & 32.67\% & 57.01\% & 6.33\% & 9.35\% & 57.00\% & 27.10\%&&\\
\hline

\end{tabular}
}
    \caption{Frequency of specific codes for the questions regarding the effects of technology and the restriction of technology on religious education courses.}
    \label{tab:codes-classes}
\end{table}

\section{Discussion}
What the data shows is that both paper scriptures and digital scriptures are part of everyday life for many religious individuals. In the following subsections, we discuss our results as they relate to each of our four research questions in turn.

\subsection{RQ1: Differences in Spiritual Experiences Across Media}
The rates at which individuals report having spiritual experiences differ depending on the medium used for scripture study, which answers research question 1. Paper scriptures had the highest frequency of spiritual experiences across both universities, and mobile devices were also common for \usu{} participants.

\churchname{} was the majority religion reported by \usu{} respondents, which may be primarily responsible for the differences between the two populations in terms of scripture study. \churchname{} has both an official mobile application, Gospel Library, and an official online library for its scriptures, both developed by the church itself. In addition to all of the religion's main scriptural texts, these contain a wide range of supplementary resources, including a concordance, a glossary of scriptural topics, lesson manuals, and recent speeches from church leaders. This could explain both the generally higher rate of usage of technology and the relative importance of accessing outside resources between \acu{} and \usu{}.

This preference for paper scriptures when it comes to spiritual experiences mirrors findings from other recent techno-spirituality research. In a mixed-methods study of Black Christian young adults aged 18--25, \citet{smith2026understanding} found that although Bible apps were used by slightly more of their survey respondents (81\%) than physical Bibles (77\%), many participants specifically turned to a physical Bible when what they wanted was to feel connected to God, describing the physical text as more tangible and personally meaningful than a digital version, in some cases because it was tied to a memento of a significant event in their faith, such as a baptism. Some participants also described intentionally avoiding their phones during in-person worship, for example by putting them on Do Not Disturb, in order to stay engaged rather than get distracted. Smith et al. argue that existing frameworks for religious technology use do not adequately capture this behavior, and propose a new ``non-tech practices'' dimension to describe the intentional choice to reject technology in favor of traditional spiritual practices. 

Our results echo this pattern in a different population: even though phones and tablets were used more often than paper scriptures for scripture study overall, our participants still turned to paper scriptures more than any other medium specifically when the goal was a spiritual experience, citing many of the same reasons, including a sense of tangible, personal connection and fewer distractions. Because the two surveys were designed independently and fielded concurrently, with ours complete before theirs was published (Section~\ref{sec:recruitment}), this agreement is an independent convergence rather than one study taking its lead from the other, which makes it stronger evidence than either result on its own. It suggests that the appeal of non-digital religious practice among younger, technology-saturated populations, and the resulting tension between digital convenience and spiritual authenticity, is not limited to a single religious tradition or demographic. It is also consistent with earlier ethnographic work: Wyche and Grinter found that their participants kept paper Bibles in private, low-traffic spaces such as bedrooms specifically to support quiet, reflective daily reading~\cite{wyche2009extraordinary}, the same kind of undistracted, personally meaningful engagement our participants associate with having a spiritual experience.

\subsection{RQ2: Media Selection by Task}
The difference in usage rates between scripture study purposes shows that individuals frequently use different media for different tasks, as hypothesized in research question 2, rather than choosing a favorite medium and sticking with it. This is supported quantitatively, as nearly every comparison of medium usage across the different study purposes was statistically significant, with family study, seeking revelation, and feeling a spiritual connection showing especially large effect sizes (Table~\ref{tab:p-methods}).

Laptops and desktops were rarely used for scripture study overall, but were relatively more common for two particular tasks: finding answers to a question, where they were used by nearly 10\% of \usu{} participants and 25\% of \acu{} participants, and preparing for a lesson or sermon, where usage rose to nearly a third of participants at both universities.

The choice between paper and phone or tablet also depended heavily on the task, though the pattern differed between the two universities. \usu{} participants were more likely to use phones and tablets than paper scriptures for most tasks, with two notable exceptions: feeling a spiritual connection and receiving personal revelation, the two tasks with the most inherent spiritual component, for which paper scriptures were preferred instead. \acu{} participants, by contrast, favored paper scriptures for nearly every task, with the exception of finding answers to questions, for which phones and tablets were slightly more common. This same task-dependence carried over into session length: phones and tablets were most often used for shorter study sessions of 15 minutes or less, while paper scriptures were more common for sessions between 15 and 45 minutes.

A few participants also reported using more than one medium during the same study session in order to combine their respective advantages, such as one respondent who described reading from paper scriptures while using a laptop alongside for additional resources and music. This task-dependent split between media also appears in prior research. Song's autoethnography of Christian prayer similarly paired a physical Bible with a digital reading plan within the same practice of Lectio Divina, deliberately drawing on both media for their respective strengths rather than treating the choice as either/or~\cite{song2025walking}, much like our participants who combined paper and digital resources within a single study session.

\subsection{RQ3: Benefits of Each Medium}
To answer research question 3, there are a number of different advantages offered by each medium. Mobile devices were favored for their convenience, ability to search for passages or topics, and their ability to access outside resources during study, while paper scriptures were valued for their ability to avoid distractions, the physical sensation of reading, and the spiritual connection individuals felt when using them. Although laptops and desktops are underutilized when compared with other media of scripture study, even they have their uses, such as searching for answers and preparing for sermons.

The distraction and tangibility advantages our participants attributed to paper have parallels elsewhere in the techno-spirituality literature. Claisse and Durrant found that phones and pop-up notifications pulled members of a Buddhist community's attention away from focused chanting during online practice, and that video-conferencing could not reproduce the sensory, embodied qualities of practicing together in person~\cite{claisse2023keeping}. Wolf et al. similarly found that online worship services lacked the experience of physical community and extraordinariness ~\cite{wolf2022spirituality}. Smith et al. likewise found that participants valued the ability to highlight and physically handle a paper Bible because it made them feel more connected to what they were reading than a digital equivalent could~\cite{smith2026understanding}, matching our own participants' description of the physical sensation of paper scriptures as a distinct benefit.

\subsection{RQ4: Technological Restrictions in Religious Education}
\label{sec:rq4}
As an answer to research question 4, it is abundantly clear from the responses that technology was a major cause of distraction in religious education classes, but removing technology altogether had some negative consequences as well. It hampered the ability of students to quickly find the passages mentioned during the class, the ability to use outside resources, and especially the ability of students to take notes during class.

For the results regarding the religious education classes, an additional factor is the difference in structure between the two universities. At \acu{}, Bible classes are part of the core curriculum, and must be taken in order to graduate. Conversely, \usu{} has no required religion classes, and many of the religion classes taken by students are not offered by the school for a grade; most of the religious education classes are instead offered by a local institution with no grades or assignments. This difference can explain the looser technology rules of \usu{} classes as well as the difference in how much taking notes is valued between the universities.

\subsection{Design Recommendations}
\label{sec:design-recommendations}
Based on our findings, we now present six design recommendations for those who would develop an application or website for reading scriptures. These recommendations come from both the strengths participants mentioned of digital scriptures, as well as attempting to emulate the strengths of paper scriptures. Firstly, to avoid ``swivel chair'' and distraction, such applications should have built-in access to outside resources, such as study helps. Second, such apps should provide robust search functionality, capable of finding specific passages based on wording or topic. Third, it should include an audio player to allow users to listen to the scriptures. Fourth, the ability to make detailed highlights, markings, and other notes would be beneficial as well. Fifth, due to the same-session shifts between paper scriptures and multiple devices that our participants reported, we suggest design for such apps support fluid continuity and hand-off ~\cite{raptis2016continuity}, which would allow users to keep their place, notes, tabs, and searches between devices. Finally, a way to combat distractions should be implemented, such as nudging the user to put the device on do not disturb mode or providing such a mode within the app itself.

Prior work also offers a caution for this kind of app design. Smith et al. found that a Bible app's gamified reading ``streak'' feature made some participants anxious and felt like an inauthentic way to track something as personal as scripture reading, and that the feature broke whenever a participant chose to read a physical Bible instead of the app~\cite{smith2026understanding}. This suggests that designers should be wary of adding engagement-driving features, such as streaks, that could make users feel judged for reading scripture on paper or that treat scripture study as a metric to optimize rather than a personal practice. On the other hand, O'Leary et al. found that the most-used features of their church health app were the ones focused specifically on scripture, such as short devotional reflections and Bible stories~\cite{oleary2022church}, suggesting that applications built specifically around scripture study, rather than religious practice more generally, may see the strongest engagement.



\section{Conclusions and Future Work}

Through this survey of 392 students at two universities, our results show that paper scriptures still hold an important role in personal study of religious texts for many, even as digital reading materials grow increasingly common. Paper was the medium with the highest reported frequency of spiritual experiences at both universities, and the frequency of spiritual experiences appears inversely related to the frequency of distraction, which was far higher for digital media than for paper (Figure~\ref{fig:likert-distr-spirit}); distraction was in turn named as a disadvantage of technology in religious education classes by nearly 88\% of \usu{} and 91\% of \acu{} respondents, the most consistent finding in our qualitative data. Notably, the persistence of paper is specific to spiritual purposes rather than a general reluctance to use technology: \usu{} participants used phones and tablets more than paper for most study purposes, but reversed that preference for exactly the two purposes with the most inherent spiritual component, feeling a spiritual connection and seeking personal revelation. Because our survey was designed independently of, and fielded concurrently with, \citet{smith2026understanding}'s study of a very different population, the agreement between our findings on this point is an independent convergence, and stronger evidence than either result alone that the deliberate choice of non-digital practice for spiritual ends is not confined to one tradition or demographic. At the same time, there is no hard line between those who use paper scriptures and those who use digital ones, as many individuals use both in different situations or even simultaneously, and while we observed a great deal of heterogeneity across populations---with \usu{} participants using technology more often than \acu{} participants. That heterogeneity is almost entirely a matter of digital use rather than paper use, with paper serving as a common baseline on top of which populations build differing amounts of digital practice. Taken together, these findings motivate the design recommendations in Section~\ref{sec:design-recommendations}: scripture applications should supply the search, external resources, and durable cross-device markings that draw people to digital media, while actively protecting the focus and sense of connection that keep people returning to paper.

Future work could be extended is to a wider population, in order to compare these findings to other Christian denominations or even other religious traditions. This would allow for a wider degree of perspectives to be heard and understood, as well as allow for the ability to contrast the usage of technology in personal study of religious texts between religions. 
We recommend that those pursuing this future work begin with discussions with religious leaders in different communities to understand the purposes for which they study their scriptures. This would enable the questions of the survey to accurately capture the experience of those of other faiths.
Another avenue of future work is through the creation and validation of a measure of the quality of study of religious texts via technology. The creation of this measure would allow for more concrete and granular experiments to be performed to learn more about the techno-spiritual practice of using technology for the personal study of religion.


\bibliographystyle{ACM-Reference-Format}
\bibliography{references}

\appendix









\section{Survey}
\label{sec:survey}
The following is a list of the survey questions asked during the study.

\begin{enumerate}
  \item For what purposes do you read or study scriptures on a regular basis (select all that apply)?
  \begin{itemize}
      \item Answers include: Prepare for a talk, lesson, sermon, etc; Personal Study; Study as a family; Study with friends/Bible study (This option was added to the \acu{} survey, \usu{} participants did not see it during the survey); Looking for an answer to a question; Seek spiritual guidance/receive personal revelation; Feel a spiritual connection; Self improvement; Other (with a text box to specify what is meant by other)
  \end{itemize}
  \item In an average week, how long do you spend on doing scripture study using each medium?
  \begin{itemize}
      \item Answers include 0 Min, 1-30 Min, 31 Min - 1 Hr, 2 Hr, and 3+ Hr for each of the media Paper, Phone/Tablet, and Laptop/Desktop Computer
  \end{itemize}
  \item How many days a week, on average, do you spend time doing scripture study with each medium?
  \begin{itemize}
      \item Answers include 0 Days, 1 Day, 2 Days, 3 Days, 4 Days, 5 Days, 6 Days, and 7 Days for each of the media Paper, Phone/Tablet, and Laptop/Desktop Computer
  \end{itemize}
  \item How often do you have a spiritual experience when using a given medium for scripture study?
  \begin{itemize}
      \item Answers include Almost never, Rarely, Sometimes, Often, and Very often for each of the media Paper, Phone/Tablet, and Laptop/Desktop Computer
  \end{itemize}
  \item How long is an average session of scripture study when you use each medium?
  \begin{itemize}
      \item Answers include Do not use this medium, 15 minutes or less, 16-30 minutes, 31-45 minutes, 46-60 minutes, and More than an hours for each of the media Paper, Phone/Tablet, and Laptop/Desktop Computer
  \end{itemize}
  
  \item (This question was only asked if the participant selected `prepare for a talk, lesson, sermon, etc' for question 1) When reading scriptures to prepare a talk (or lesson, sermon, etc), how often do you use each medium?
  \begin{itemize}
      \item Answers include Almost never, Rarely, Sometimes, Often, and Very often for each of the media Paper, Phone/Tablet, and Laptop/Desktop Computer
  \end{itemize}
  \item (This question was only asked if the participant selected `prepare for a talk, lesson, sermon, etc' for question 1) Explain why you prefer the medium you do for preparing a talk/lesson/sermon
  \begin{itemize}
      \item Answer was a free response text box
  \end{itemize}
  
  \item (This question was only asked if the participant selected `personal study' for question 1) When reading scriptures for personal study, how often do you use each medium?
  \begin{itemize}
      \item Answers include Almost never, Rarely, Sometimes, Often, and Very often for each of the media Paper, Phone/Tablet, and Laptop/Desktop Computer
  \end{itemize}
  \item (This question was only asked if the participant selected `personal study' for question 1) Explain why you prefer the medium you do for personal study
  \begin{itemize}
      \item Answer was a free response text box
  \end{itemize}
  
  \item (This question was only asked if the participant selected `study as a family' for question 1) When reading scriptures for family study, how often do you use each medium?
  \begin{itemize}
      \item Answers include Almost never, Rarely, Sometimes, Often, and Very often for each of the media Paper, Phone/Tablet, and Laptop/Desktop Computer
  \end{itemize}
  \item (This question was only asked if the participant selected `study as a family' for question 1) Explain why you prefer the medium you do for family study
  \begin{itemize}
      \item Answer was a free response text box
  \end{itemize}
  
  \item (This question was only asked if the participant selected `study with friends/Bible study' for question 1) When reading scriptures for studying with friends/Bible study, how often do you use each medium?
  \begin{itemize}
      \item Answers include Almost never, Rarely, Sometimes, Often, and Very often for each of the media Paper, Phone/Tablet, and Laptop/Desktop Computer
  \end{itemize}
  \item (This question was only asked if the participant selected `study with friends/Bible study' for question 1) Explain why you prefer the medium you do for studying with friends/Bible study
  \begin{itemize}
      \item Answer was a free response text box
  \end{itemize}
  
  \item (This question was only asked if the participant selected `looking for an answer to a question' for question 1) When reading scriptures to look for answers to questions, how often do you use each medium?
  \begin{itemize}
      \item Answers include Almost never, Rarely, Sometimes, Often, and Very often for each of the media Paper, Phone/Tablet, and Laptop/Desktop Computer
  \end{itemize}
  \item (This question was only asked if the participant selected `looking for an answer to a question' for question 1) Explain why you prefer the medium you do for looking for an answer to a question
  \begin{itemize}
      \item Answer was a free response text box
  \end{itemize}
  
  \item (This question was only asked if the participant selected `seek spiritual guidance/receive personal revelation' for question 1) When reading scriptures to seek spiritual guidance or receive personal revelation, how often do you use each medium?
  \begin{itemize}
      \item Answers include Almost never, Rarely, Sometimes, Often, and Very often for each of the media Paper, Phone/Tablet, and Laptop/Desktop Computer
  \end{itemize}
  \item (This question was only asked if the participant selected `seek spiritual guidance/receive personal revelation' for question 1) Explain why you prefer the medium you do for seeking spiritual guidance/receiving personal revelation
  \begin{itemize}
      \item Answer was a free response text box
  \end{itemize}

  \item (This question was only asked if the participant selected `feel a spiritual connection' for question 1) When reading scriptures to feel a spiritual connection, how often do you use each medium?
  \begin{itemize}
      \item Answers include Almost never, Rarely, Sometimes, Often, and Very often for each of the media Paper, Phone/Tablet, and Laptop/Desktop Computer
  \end{itemize}
  \item (This question was only asked if the participant selected `feel a spiritual connection' for question 1) Explain why you prefer the medium you do for feeling a spiritual connection
  \begin{itemize}
      \item Answer was a free response text box
  \end{itemize}

  \item (This question was only asked if the participant selected `self improvement' for question 1) When reading scriptures for self improvement, how often do you use each medium?
  \begin{itemize}
      \item Answers include Almost never, Rarely, Sometimes, Often, and Very often for each of the media Paper, Phone/Tablet, and Laptop/Desktop Computer
  \end{itemize}
  \item (This question was only asked if the participant selected `self improvement' for question 1) Explain why you prefer the medium you do for self improvement
  \begin{itemize}
      \item Answer was a free response text box
  \end{itemize}

  \item (This question was only asked if the participant selected `other' for question 1) When reading scriptures for [REASON ENTERED IN `OTHER' TEXT BOX], how often do you use each medium?
  \begin{itemize}
      \item Answers include Almost never, Rarely, Sometimes, Often, and Very often for each of the media Paper, Phone/Tablet, and Laptop/Desktop Computer
  \end{itemize}
  \item (This question was only asked if the participant selected `other' for question 1) Explain why you prefer the medium you do for reading for [REASON ENTERED IN `OTHER' TEXT BOX]
  \begin{itemize}
      \item Answer was a free response text box
  \end{itemize}

  \item In which situations not yet mentioned do you prefer phone/tablet scriptures, if at all?
  \begin{itemize}
      \item Answer was a free response text box
  \end{itemize}
  \item In which situations not yet mentioned do you prefer laptop/desktop scriptures, if at all?
  \begin{itemize}
      \item Answer was a free response text box
  \end{itemize}
  \item In which situations not yet mentioned do you prefer paper scriptures, if at all?
  \begin{itemize}
      \item Answer was a free response text box
  \end{itemize}
  \item How often do you experience distraction with each of the following media?
  \begin{itemize}
      \item Answers include Almost never, Rarely, Sometimes, Often, and Very often for each of the media Paper, Phone/Tablet, and Laptop/Desktop Computer
  \end{itemize}

  \item How often has your use of technology disrupted your learning experience for religious education classes?
  \begin{itemize}
      \item Answers include Not applicable, Almost Never, Rarely, Sometimes, Often and Very often
  \end{itemize}
  \item How often has someone else's use of technology disrupted your learning experience for religious education classes?
  \begin{itemize}
      \item Answers include Not applicable, Almost Never, Rarely, Sometimes, Often and Very often
  \end{itemize}
  \item What advantages have you seen, if any, from the use of technology in religious education classes?
  \begin{itemize}
      \item Answer was a free response text box
  \end{itemize}
  \item What disadvantages have you seen, if any, from the use of technology in religious education classes?
  \begin{itemize}
      \item Answer was a free response text box
  \end{itemize}
  \item If you have participated in a religious education class where technology was restricted, what kind of class was it?
  \begin{itemize}
      \item Answer was a free response text box
  \end{itemize}
  \item If you have participated in a religious education class where technology was restricted, what restrictions were put in place?
  \begin{itemize}
      \item Answer was a free response text box
  \end{itemize}
  \item If you have participated in a religious education class where technology was restricted, how did those restrictions impact your learning experience?
  \begin{itemize}
      \item Answer was a free response text box
  \end{itemize}
  \item Is there anything else you'd like to share regarding your experience with religion and technology?
  \begin{itemize}
      \item Answer was a free response text box
  \end{itemize}
  
  \item How would you describe your gender?
  \begin{itemize}
      \item Answers include Man, Woman, Non-binary, Prefer not to say, and Other (with a text box to specify)
  \end{itemize}
  \item What is your ethnicity?
  \begin{itemize}
      \item Answers include Native American, Asian / Pacific Islander, Black or African American, Hispanic, White / Caucasian, Prefer not to say, and Other (with a text box to specify). Multiple ethnicities can be selected. 
  \end{itemize}
  \item What is your age?
  \begin{itemize}
      \item Answers include 18-22, 23-25, 26-34, 35-44, 45-54, 55-64, Above 64, and Prefer not to say.
  \end{itemize}
  \item What is your religious affiliation? 
  \begin{itemize}
      \item Answers include Catholic, Church of Jesus Christ of Latter-Day Saints, Jehovah's Witness, Churches of Christ, Baptist, Other Christian Denomination, Jewish, Muslim, Hindu, Buddhist, Other (with a text box to specify), and Prefer not to say
  \end{itemize}
  \item What is your major? (Enter `None' if not applicable)
  \begin{itemize}
      \item Answers were in a drop down box containing a list of majors, along with the options of None, Prefer not to say, and Other.
  \end{itemize}
  \item (This question was only asked if the answer to the previous was `Other') `Other' was selected as your major. What is your major?
  \begin{itemize}
      \item Answer was a free response text box.
  \end{itemize}
\end{enumerate}

\end{document}